\documentclass[review]{elsarticle}

\usepackage{lineno,hyperref}
\modulolinenumbers[5]
\usepackage{graphicx}
\usepackage{amsmath, amssymb}
\usepackage{tikz}
\usepackage{upgreek}
\biboptions{sort&compress}
\graphicspath{{./fig/}}

\usepackage{color}
\usepackage{xcolor}

\colorlet{rebuttal}{black!62!green}

\newcommand{\Pran}{\text{\textit{Pr}}}
\newcommand{\Gras}{\text{\textit{Gr}}}

\newcommand{\ord}{\text{\textit{O}}}

\newcommand{\ie}{i.e.\ }
\newcommand{\etal}{\text{\textit{et al.\ }}}

\newcommand{\p}{\partial}

\newcommand{\vect}[1]{\boldsymbol{#1}}

\newcommand{\vv}{\vect{v}}
\newcommand{\VV}{\vect{\bar{v}}}

\newcommand{\betas}{\beta^*}
\newcommand{\ts}{t^*}
\newcommand{\xs}{x^*}

\newcommand{\zs}{z^*}
\newcommand{\us}{u^*}
\newcommand{\vs}{v^*}
\newcommand{\ws}{w^*}
\newcommand{\gs}{g^*}
\newcommand{\nablas}{\nabla^*}
\newcommand{\rhos}{\rho^*}
\newcommand{\mus}{\mu^*}
\newcommand{\nus}{\nu^*}
\newcommand{\kappas}{\kappa^*}
\newcommand{\Ts}{T^*}
\newcommand{\ps}{p^*}
\newcommand{\vvs}{\vv^*}
\newcommand{\deltas}{\delta^*_o}

\newcommand{\uh}{\hat{u}}
\newcommand{\vh}{\hat{v}}
\newcommand{\wh}{\hat{w}}
\newcommand{\ph}{\hat{p}}
\newcommand{\thetah}{\hat{\theta}}
\newcommand{\rhoh}{\hat{\rho}}
\newcommand{\muh}{\hat{\mu}}

\newcommand{\ut}{\tilde{u}}
\newcommand{\vt}{\tilde{v}}
\newcommand{\thetat}{\tilde{\theta}}

\newcommand{\Grast}{\tilde{\Gras}}

\newcommand{\Psis}{\psi^*}

\newcommand{\Uc}{u_o^*}

\newcommand{\ue}{\mathrm{e}}
\newcommand{\ui}{\mathrm{i}}

\newcommand{\omi}{\omega_{\text{i}}}
\newcommand{\omr}{\omega_{\text{r}}}

\newcommand{\ms}{\kern.10em\relax}

\newcommand{\pfr}[2]{\ensuremath{\frac{\partial #1}{\partial #2}}}

\journal{Journal of \LaTeX\ Templates}

\begin{document}

\begin{frontmatter}

\title{Non-Boussinesq stability analysis of natural-convection gaseous flow on inclined hot plates}

\author[mainaddress]{Prabakaran Rajamanickam\corref{mycorrespondingauthor}} 
\cortext[mycorrespondingauthor]{Corresponding author}
\ead{prajaman@ucsd.edu}

\author[mainaddress]{Wilfried Coenen} 
\author[mainaddress]{Antonio L. S\'anchez}
\address[mainaddress]{Department of Mechanical and Aerospace Engineering, University of California San Diego, La Jolla, CA 92093--0411, USA}



\begin{abstract}
The buoyancy-driven boundary-layer flow that develops over a semi-infinite inclined hot plate is
known to become unstable at a finite distance from the leading edge, characterized by a critical
value of the Grashof number $\Gras$ based on the local boundary-layer thickness. The nature
of the resulting instability depends on the inclination angle $\phi$, measured from the vertical
direction. For values of $\phi$ below a critical value $\phi_c$ the instability is characterized
by the appearance of spanwise traveling waves, whereas for $\phi>\phi_c$ the bifurcated flow
displays G\"ortler-like streamwise vortices. The Boussinesq approximation, employed in previous
linear stability analyses, ceases to be valid for gaseous flow when the wall-to-ambient
temperature ratio $\Theta_w$ is not close to unity. The corresponding non-Boussinesq analysis is
presented here, accounting also for the variation with temperature of the different transport
properties. A temporal stability analysis including nonparallel effects of the base flow is used
to determine curves of neutral stability, which are then employed to delineate the dependences
of the critical Grashof number and of its associated wave length on the inclination angle $\phi$
and on the temperature ratio $\Theta_w$ for the two instability modes, giving quantitative
information of interest for configurations with $\Theta_w-1\sim 1$. The analysis provides in
particular the predicted dependence of the crossover inclination angle $\phi_c$ on $\Theta_w$,
indicating that for gaseous flow with $\Theta_w-1\sim 1$ spanwise traveling waves are
predominant over a range of inclination angles $0 \le \phi \le \phi_c$ that is significantly
wider than that predicted in the Boussinesq approximation.
\end{abstract}


\end{frontmatter}


\section{Introduction}
\label{sec:intro}

A semi-infinite inclined hot plate placed in a quiescent air atmosphere is known to induce near
its surface a free-convection flow as a result of the action of buoyancy forces on the heated
gas. The structure of the resulting boundary layer away from the plate edge exhibits at leading
order a self-similar structure, as first noted in the experimental study of Schmidt and Beckmann
\cite{SB1930}. This boundary layer is known to become unstable to small disturbances at a
certain distance measured from the leading edge of the plate \cite{Sparrow1969}. The character
of the observed instability depends on the inclination angle $\phi$, measured from the vertical
direction. Thus, for values of $\phi$ above a critical value $\phi_c$, including in particular
horizontal and nearly horizontal plates, the instability develops in the form of stationary
counter-rotating vortex rolls that are oriented in the streamwise direction. These are similar
to those characterizing the G\"ortler instability of boundary-layer flow along a concave wall,
driven by centrifugal forces, with the wall-normal component of the buoyancy force being the
driving mechanism for free-convection flow. As the inclination angle $\phi$ is decreased, this
wall-normal buoyancy component loses importance and, below a certain crossover angle $\phi_c$,
the character of the observed instabilities changes to Tollmien-Schlichting-like traveling waves
driven by shear. Following existing terminology \cite{Haaland1973, HaalandS1973}, in the
following the stability mode involving streamwise stationary vortices will be termed
\emph{vortex instability}, whereas that involving traveling waves will be termed \emph{wave
instability}.

Sparrow \& Husar~\cite{Sparrow1969} were the first to identify both modes experimentally, and to
show that their prevalence depends on the inclination of the heated surface. The crossover angle
was determined by Lloyd \& Sparrow~\cite{Lloyd1970} to lie between $14^{\circ} < \phi_c <
17^{\circ}$. Other experiments carried out later agree generally with these findings
\cite{Gebhart1978, Cheng1988, Zuercher1998, Jeschke2000, Trautman2002, Kimura2003}.

Apart from the inclination angle $\phi$, the buoyancy-induced flow over a semi-infinite flat
plate at constant temperature depends on the Prandtl number $\Pran$ of the fluid and on the
ratio $\Theta_w = \Ts_w / \Ts_\infty$ of the wall temperature to the ambient temperature. All
previous theoretical efforts aimed at quantifying the critical conditions at the onset of the
vortex and wave instabilities were performed in the Boussinesq approximation \cite{Haaland1973,
HaalandS1973, Hwang1973, Kahawita1974, Iyer1974, Chen1982, Tzuoo1985, Lin2001, Tien1986,
Chen1991, Tumin2003}, which is only justified in gaseous flow when the wall-to-ambient
relative temperature difference $(\Theta_w-1)$ is small. Most of these studies employ linear
local stability theory---a normal mode analysis---to determine, for fixed values of $\Pran$ and
$\phi$, the critical boundary-layer thickness $\deltas$, measured in dimensionless form through
a local Grashof number, above which small perturbations, either of vortex type with associated
spanwise wave number $l^*$, or of wave type with streamwise wave number $k^*$, are amplified. In
this manner, a unique neutral curve in the Grashof -- wave number plane can be delineated for
each mode. The mode with the lowest corresponding critical Grashof number for all wave numbers
would be the one that prevails in an experiment, and the value of that Grashof number would give
the local boundary-layer thickness---and therefore the distance $\xs$ to the plate edge---at
which the instability first develops.

Conventionally, in a local stability analysis the base flow is assumed to be strictly parallel.
That assumption must be reconsidered in the analysis of slowly varying slender flows, such as
the present boundary layer, for which the order of magnitude of some of the terms in the
stability equations, involving the transverse velocity component and the streamwise variation of
the flow, is comparable to that of the viscous terms, and must be correspondingly taken into
account. This so-called locally nonparallel approach was already adopted by Haaland \& Sparrow
in their temporal stability analyses of the vortex~\cite{Haaland1973} and
wave~\cite{HaalandS1973} instability modes. The resulting critical Grashof numbers were seen to
differ by several orders of magnitude from those obtained with a strictly parallel analysis,
thereby underlining the importance of the nonparallel terms. The problem was re-examined by a
series of authors, adopting small variations of this approach, either in a temporal
\cite{Hwang1973, Kahawita1974, Chen1982, Tzuoo1985, Lin2001, Tien1986, Chen1991} or a spatial
\cite{Iyer1974} linear--stability framework. The analysis can be extended to describe
finite-amplitude vortex rolls and secondary bifurcations by retaining selected nonlinear terms
in the description, as done by Chen~\etal\cite{Chen1991}. Recently,
instabilities in transient cooling of inclined surfaces and cavities have been
studied numerically~\cite{Saha2011, Dou2013}.

The accuracy of the computations mentioned above deteriorates in the presence of order--unity
deviations of the wall temperature from the ambient temperature, when the use of the Boussinesq
approximation is no longer justified. Although non-Boussinesq effects have been taken into
account in analyses of the boundary-layer structure for flow over a heated plate
\cite{Clarke1975, Ackroyd1976, Sanchez2013}, these effects have never been considered in
connection with the associated stability problem. The objective of the present work is to
revisit the classical work of Haaland \& Sparrow \cite{Haaland1973, HaalandS1973}, including the
influence of the wall-to-ambient temperature ratio $\Theta_w$ for cases with $\Theta_w-1 \sim
1$. In particular, a linear temporal modal stability analysis will be performed to investigate
both the vortex and the wave modes, with account taken of nonparallel effects associated with
the slow downstream evolution of the base flow. The effect of $\Theta_w$ on the neutral
stability curves will be assessed, along with the dependence of the crossover angle defining the
transition between the two types of instability.

The paper is structured as follows. The governing equations and boundary conditions for the base
flow and for the linear stability analysis are given in \S\,2. The vortex mode is studied in
\S\,3, followed in \S\,4 by the analysis of the wave instability. The predictions of the
critical conditions for the two modes are compared in \S\,5 to delineate the boundary that
defines the regions of prevalence of each mode on the parametric plane $\phi-\Theta_w$.
Finally, concluding remarks are offered in \S\,6.

\section{Problem formulation}
\label{sec:form}

\begin{figure}
\centering
\includegraphics[scale=0.6]{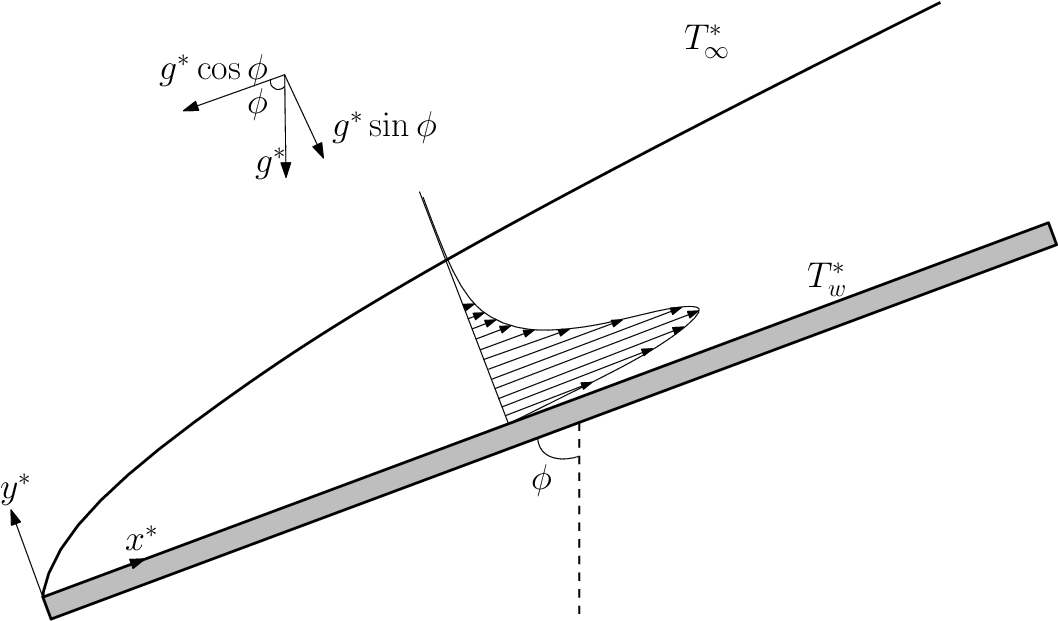}
\caption{Schematic diagram of the boundary-layer flow over a heated inclined surface.}
\label{fig:scheme}
\end{figure}

The problem considered here, shown schematically in figure~\ref{fig:scheme}, involves the flow
induced by buoyancy near the surface of a semi-infinite inclined plate whose temperature is held
at a constant value, $\Ts_w$, higher than the ambient temperature $\Ts_\infty$ found in the
surrounding quiescent air atmosphere. The associated velocities are negligibly small compared to
the sound speed, so that the conservation equations can be written in the low-Mach number
approximation
\begin{align}
    \pfr{\rhos}{\ts} + \nablas \cdot (\rhos\vvs) &= 0, \label{eq:nsdim1} \\
    \rhos\pfr{\vvs}{\ts} + \rhos\vvs\cdot\nablas\vvs &=
        - \nablas \ps
        + (\rhos-\rhos_\infty) \vect{\gs}
        + \nablas \cdot \left[ \mus (\nablas \vvs + {\nablas \vvs}^T) \right], \label{eq:nsdim2} \\
    \rhos \pfr{\Ts}{\ts} + \rhos \vvs\cdot\nablas \Ts &=
          \frac{1}{\Pran} \nablas \cdot (\mus \nablas \Ts), \label{eq:nsdim3}
\end{align}
where $\rhos$, $\vvs$, and $\Ts$ represent the density, velocity, and temperature of the gas; dimensional quantities are indicated everywhere in the text with an asterisk $({}^*)$. In the momentum equation~\eqref{eq:nsdim2}, $\ps$ represents the sum of the pressure difference from the ambient hydrostatic distribution and the isotropic component of the stress tensor. Cartesian coordinates are used in the description, including the streamwise distance measured along the plate from the leading edge $\xs$, the transverse distance from the surface of the plate $y^*$, and the spanwise coordinate $\zs$, with corresponding velocity components $\vvs = (\us, \vs, \ws)$. The inclination angle $\phi$ is measured from the vertical, so that the gravity vector is $\vect{\gs} = -\gs \cos\phi \, \vect{e_x} - \gs \sin\phi \, \vect{e_y}$.

The above equations must be supplemented with the equation of state
\begin{equation}
     \frac{\rhos}{\rhos_\infty} = \frac{\Ts_\infty}{\Ts}
\end{equation}
and with the presumed power law 
\begin{equation}
     \frac{\mus}{\mus_\infty} = \frac{\kappas}{\kappas_\infty} = \left( \frac{\Ts}{\Ts_\infty} \right)^\sigma \label{sigma}
\end{equation}
for the temperature dependence of the viscosity and thermal conductivity, with the subscript $\infty$ denoting properties in the unperturbed ambient air. The constant values $\Pran = 0.7$ and $\sigma = 2/3$, corresponding to air, will be used below for the Prandtl number $\Pran = c_p^* \mus_\infty/\kappas_\infty$ in~\eqref{eq:nsdim3} and for the exponent $\sigma$ in~\eqref{sigma}. Equations~\eqref{eq:nsdim1}--\eqref{eq:nsdim3} must be integrated with the boundary conditions
\begin{equation}
\left\{ \begin{array}{ll} \us=\vs=\ws=\Ts-\Ts_w=0 & {\rm at} \quad y^*=0 \quad {\rm for} \quad \xs>0 \\
\us=\vs=\ws=\Ts-\Ts_\infty=\ps=0 & {\rm as} \quad (x^{*2}+y^{*2}) \rightarrow \infty \quad {\rm for}  \quad \; y^* \ne 0, \xs>0.
\end{array} \right.
\label{bc}
\end{equation}

For plates that are not nearly horizontal, such that $\pi/2-\phi$ is not small, the flow is driven by the direct acceleration associated with the gravity component parallel to the plate $\gs \cos\phi$. Near the leading edge of the plate there exists a nonslender Navier--Stokes region of characteristic size $[{\nus_\infty}^2/(\gs \cos\phi)]^{1/3}$ where the velocity components are of order $({\nus_\infty} \gs \cos\phi)^{1/3}$, with $\nus_\infty=\mus_\infty/\rhos_\infty$ denoting the ambient kinematic viscosity, such that the local Reynolds number there is of order unity. Outside this Navier--Stokes region the flow--field structure includes a boundary-layer region of characteristic thickness $[({\nus_\infty}^2 \xs)/(\gs \cos\phi)]^{1/4}$ and characteristic streamwise velocity $\left(\gs \cos\phi \, \xs\right)^{1/2}$, surrounded by an outer region of slow irrotational motion driven by the boundary-layer entrainment. 

The stability of the boundary layer at a given location $\xs=\xs_o$ depends on the value of the associated Grashof number
\begin{equation}
    \Gras = \frac{\xs_o}{\deltas}=\frac{{\deltas}^3 \gs \cos\phi}{{\nus_\infty}^2}= \left( \frac{{\xs_o}^3 \gs \cos\phi}{{\nus_\infty}^2} \right)^{1/4}, \label{Grasdef}
\end{equation}
which is the Reynolds number based on the local values of the thickness and streamwise velocity 
\begin{equation}
\deltas= \left(\frac{{\nus_\infty}^2 \xs_o}{\gs \cos\phi} \right)^{1/4} \quad {\rm and} \quad \Uc=\left(\gs \cos\phi \, \xs_o \right)^{1/2}.
\end{equation}
The following analysis assumes implicitly that the critical value of $\Gras$ associated with the
onset of the instability is moderately large, as corresponds to a location $\xs_o/\deltas \gg 1$
far downstream from the Navier--Stokes region, where the flow near the plate surface is slender,
enabling the stability analysis to be developed on the basis of the nearly parallel approximation, with the self-similar boundary-layer solution used to evaluate the base flow, as indicated below.

\subsection{Base flow}

The near-plate solution that develops outside the Navier--Stokes region can be described with small relative errors of order $\Gras^{-2}$ by using the boundary-layer form of the conservation equations. We shall neglect the pressure differences across the boundary layer, of order $\rho^*_\infty g^* \sin \phi \, \delta^*_o$, associated with the transverse component of the gravitational acceleration $g^* \sin \phi$, because they are small compared with the streamwise hydrostatic pressure differences $\rho^*_\infty g^* \cos \phi \, x^*_o$ at distances $x^*_o \gg \delta^*_o \tan \phi$, corresponding to local values of the Grashof number 
\begin{equation}
\Gras \gg \tan\phi,  \label{condition}
\end{equation}
a condition readily satisfied outside the Navier--Stokes region for the inclined plates with
$\pi/2-\phi \sim 1$ considered here. The resulting solution is self-similar when expressed in terms of the dimensionless coordinate 
\begin{equation}
\eta = \frac{y^*}{[({\nus_\infty}^2 \xs)/(\gs \cos\phi)]^{1/4}}
\end{equation}
with use made of the nondimensional temperature and stream function
 \begin{equation}
 \Theta(\eta) = \frac{\Ts}{\Ts_\infty} \quad {\rm and} \quad F(\eta) = \frac{\Psis}{\left(x^{*3} {\nus_\infty}^2 \gs \cos\phi \right)^{1/4}},
\end{equation}
with associated self-similar velocity components
 \begin{equation}
U=\frac{u^*}{(\gs \cos\phi \, \xs)^{1/2}}=\Theta F' \quad \text{and} \quad V=\frac{v^*}{({\nus_\infty}^2 \gs \cos\phi / \xs)^{1/4}}=\frac{\Theta}{4} \left( \eta F' - 3 F \right).
\end{equation}
Here the prime is used to denote differentiation with respect to $\eta$. Introducing these variables reduces the problem to that of integrating
\begin{align}
    (\Theta^\sigma(\Theta F')')'
        + \frac{3}{4}F(\Theta F')'
        - \frac{1}{2}\Theta F'^2
        + 1
        - \frac{1}{\Theta}
        & = 0, \label{eq:selfsimmom}\\
    (\Theta^\sigma \Theta')'
        + \frac{3}{4}\Pran F \Theta'
        & = 0, \label{eq:selfsimener}
\end{align}
with boundary conditions $F(0) = F'(0) = F'(\infty) = \Theta(0) - \Theta_w = \Theta(\infty) - 1 = 0$, where $\Theta_w = \Ts_w/\Ts_\infty$ is the wall-to-ambient temperature ratio. 

\begin{figure}
\centering
\includegraphics[scale=0.5]{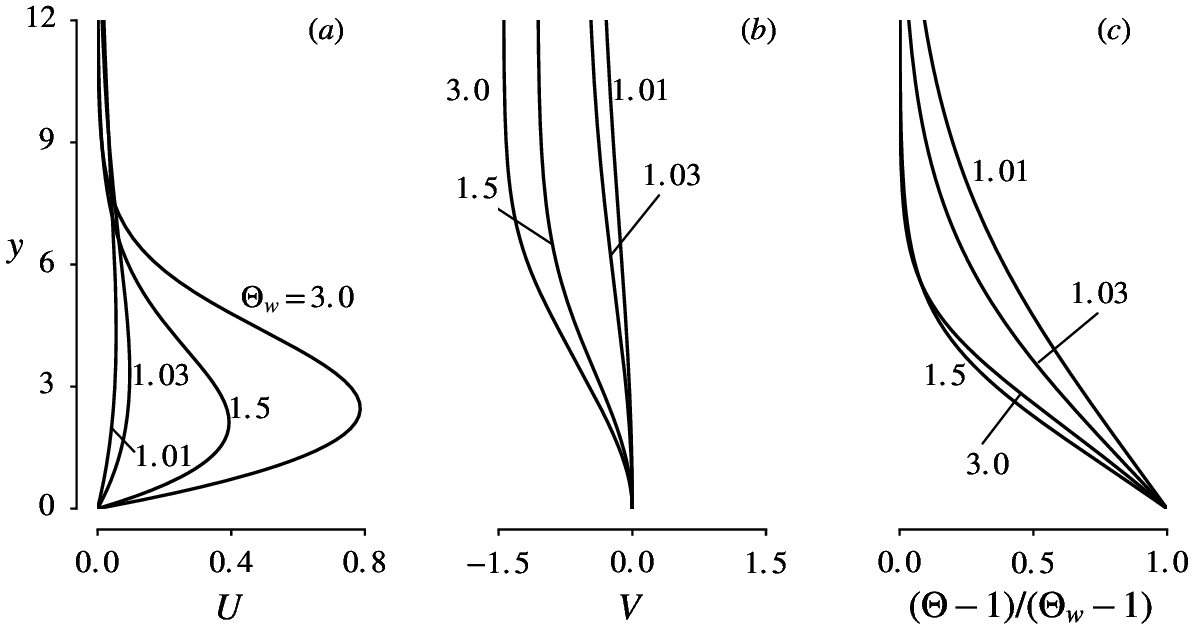}
\caption{The self-similar base-flow solution obtained by integration of~\eqref{eq:selfsimmom}--\eqref{eq:selfsimener} with $\Pran = 0.7$ and $\sigma = 2/3$ for different values of $\Theta_w$.
}
\label{fig:baseflow}
\end{figure}

The numerical integration was performed using a spectral collocation method with a fractional
transformation defined by $y=a(1+\xi)/(b-\xi)$ \cite{schmid2012}, where $a = y_c \,
y_\text{max}/(y_\text{max}-2y_c)$ and $b=1+2a/y_\text{max}$, that maps $N$ Chebyshev collocation
points $\xi\in[-1,1]$ to the physical domain $y\in[0,y_\text{max}]$ allocating $N/2$ collocation
points to the near-wall region $0<y<y_c$. Values $y_\text{max}=200$, $y_c=15$ and $N=128$ are
found to be adequate, slightly higher than the values used in \cite{Tumin2003} for spatial
analysis. Resulting temperature and velocity profiles are shown in figure~\ref{fig:baseflow} for
different values of $\Theta_w$, giving results similar to those presented elsewhere
\cite{Sanchez2013}.

\subsection{Linear stability analysis}

Using standard practice, the linear stability of the flow at $\xs=\xs_o$ is investigated by introducing into~\eqref{eq:nsdim1}--\eqref{sigma} the perturbed nondimensional variables
\begin{align}
\frac{\vvs}{\Uc}&=\VV(x/\Gras,y)+\vv(x,y,z,t)=(\bar{u},\bar{v},0)+(u,v,w) \nonumber \\
\frac{T^*}{T^*_\infty}&=\overline{T}(x/\Gras,y)+\theta(x,y,z,t) \nonumber  \\
\frac{\rho^*}{\rho^*_\infty}&=\bar{\rho}(x/\Gras,y)+\rho(x,y,z,t) \label{decomposition} \\
\frac{\mu^*}{\mu^*_\infty}&=\bar{\mu}(x/\Gras,y)+\mu(x,y,z,t) \nonumber \\
\frac{p^*}{\rho^*_\infty {\Uc}^2}&=p(x,y,z,t) \nonumber 
\end{align}
where
\begin{equation}
x=\frac{x^*-x^*_o}{\deltas}, \; y=\frac{y^*}{\deltas}, \; z=\frac{z^*}{\deltas}, \quad {\rm and}
\quad t=\frac{t^*}{\deltas/\Uc},
\end{equation}
leading to the linearized equations 
\begin{align}
    \pfr{\rho}{t} &= - \bar{\rho} \, \nabla\cdot\vv - \rho \, \nabla\cdot\VV - \vv \cdot \nabla \bar{\rho} - \VV \cdot \nabla \rho, \label{eq:lin1}\\
    \bar{\rho} \, \pfr{\vv}{t} &=
        - \bar{\rho} \, \VV \cdot \nabla \vv
        - \bar{\rho} \, \vv \cdot \nabla \VV
        - \rho \, \VV \cdot \nabla \VV
        - \nabla p
        - \frac{1}{\Gras} \rho (\vect{e_x} + \tan \phi \, \vect{e_y}) \nonumber \\
        & \quad + \frac{1}{\Gras} \nabla \cdot \left[   \bar{\mu} (\nabla \vv + {\nabla \vv}^T)
                                            + \mu (\nabla \VV + {\nabla \VV}^T) \right], \label{eq:lin2}\\
    \bar{\rho} \, \pfr{\theta}{t} & =
        - \bar{\rho} \, \VV \cdot \nabla \theta
        - \bar{\rho} \, \vv \cdot \nabla \overline{T}
        - \rho \, \VV \cdot \nabla \overline{T} \nonumber \\
        & \quad + \frac{1}{\Gras\,\Pran} \nabla \cdot \left[ \bar{\mu} \nabla \theta + \mu \nabla \overline{T} \right], \label{eq:lin3}
    \intertext{and}
    \rho & = -\overline{T}^{-2} \theta
    \qquad \text{and} \qquad
    \mu = \sigma \overline{T}^{\sigma-1} \theta. \label{eq:lin4}
\end{align}

The base profiles $\VV$, $\overline{T}$, $\bar{\rho}$, and $\bar{\mu}$ can be evaluated from the
self-similar velocity and temperature profiles $U(\eta)$, $V(\eta)$, and $\Theta(\eta)$ according to
\begin{align}
    &\bar{u} = (1+x/\Gras)^{1/2} \, U(y/[1+x/\Gras]^{1/4}), \nonumber \\ 
    &\bar{v} = \Gras^{-1}  (1+x/\Gras)^{-1/4} \, V({y}/{[1+x/\Gras]^{1/4}}), \nonumber \\
    &\overline{T} = \frac{1}{\bar{\rho}}=\bar{\mu}^{1/\sigma}=\Theta({y}/{[1+x/\Gras]^{1/4}}),
\end{align}
written in terms of the local coordinates $x=(x^*-x^*_o)/\delta^*_o$ and $y=y^*/\delta^*_o$ by using $\eta=y/(1+x/\Gras)^{1/4}$ and $\Gras=x^*_o/\delta^*_o$. As can be seen, the base flow displays a slow streamwise variation through the rescaled coordinate $x/\Gras$. The expressions given above can be used to evaluate the factors that appear in the linearized equations~\eqref{eq:lin1}--\eqref{eq:lin3}. In the first approximation for $\Gras \gg 1$ one obtains the order-unity factors
\begin{align}
&\bar{u}=U(y), \frac{\p \bar{u}}{\p y}=U'(y), \frac{\p^2 \bar{u}}{\p y^2}=U''(y), \nonumber \\ & \overline{T}=\frac{1}{\bar{\rho}}=\bar{\mu}^{1/\sigma}=\Theta(y), 
\frac{\p \overline{T}}{\p y}=\Theta'(y), \frac{\p^2 \overline{T}}{\p y^2}=\Theta''(y), 
\end{align}
which pertain to the strictly parallel flow, along with the factors of order $\Gras^{-1}$
\begin{align}
&\bar{v}= \Gras^{-1} V(y), \; \frac{\p \bar{v}}{\p y}=\Gras^{-1} V'(y), \frac{\p \bar{u}}{\p x}=\Gras^{-1} \left[\frac{1}{2} U(y) - \frac{y}{4} U'(y)\right], \nonumber \\
&\frac{\p \overline{T}}{\p x}=-\Gras^{-1} \frac{y}{4} \Theta'(y), \quad \text{and} \quad \frac{\p \bar{\rho}}{\p x}= \Gras^{-1} \frac{y}{4} \frac{\Theta'(y)}{\Theta^2(y)}, \label{factorsnonparallel}
\end{align}
arising from nonparallel effects. 

The nearly parallel stability analysis performed here assumes implicitly a moderately small value of $\Gras^{-1}$, a condition needed to ensure the slenderness of the flow. The small parameter $\Gras^{-1}$ appears in~\eqref{eq:lin1}--\eqref{eq:lin3} directly as a factor in the molecular transport terms in~\eqref{eq:lin2} and~\eqref{eq:lin3} and also through the factors~\eqref{factorsnonparallel} associated with the slow streamwise variation of the base flow. If only terms of order unity are retained in~\eqref{eq:lin1}--\eqref{eq:lin3}, then the resulting equations describe the inviscid instability of strictly parallel flow. By retaining also terms of order  $\Gras^{-1}$ one accounts simultaneously for viscous and nonparallel effects, that being the approach pursued herein. Thus, following Haaland \& Sparrow~\cite{Haaland1973, HaalandS1973}, the equations~\eqref{eq:lin1}--\eqref{eq:lin3} were systematically simplified by retaining terms up to $\ord(\Gras^{-1})$, while neglecting terms of order $\Gras^{-2}$ and higher. The resulting simplified equations are written in \ref{sec:appstabeqs} for the normal-mode perturbations 
\begin{equation}
     [u,v,w,p,\theta,\rho,\mu](x,y,z,t) =\ue^{\ui(k x + l z - \omega t)} [\uh, \vh, \wh, \ph, \thetah, \rhoh, \muh](y),
     \label{eq:normalmodes}
\end{equation}
where $k$ and $l$ are the dimensionless streamwise and spanwise wave numbers, and $\omega = \omr + \ui \, \omi$ contains both the frequency $\omr$ and the growth rate $\omi$ of the perturbations. The vortex instability is associated with modes with $k = 0$ and $\omr = 0$, whereas the wave instability corresponds to $l = 0$ and $\omr > 0$. The two types of instabilities will be studied separately in the following two sections.

\section{Vortex instability}
\label{sec:gortler}

\subsection{The simplified eigenvalue problem}

We start by considering instabilities characterized by the appearance of G\"ortler-like vortex rolls aligned with the streamwise direction, known to dominate the boundary-layer dynamics for values of $\phi$ above a critical value $\phi_c$. These can be investigated by setting $k = 0$ in the normal-mode ansatz~\eqref{eq:normalmodes}. Chen~\etal\cite{Chen1991} showed that the principle of exchange of instabilities holds for this type of instabilities, in the sense that in the full eigenvalue spectrum the eigenvalue with the largest growth rate $\omega_i$ has zero frequency $\omr = 0$. Hence, the critical conditions for vortex instability can be computed by setting both $k$ and $\omega$ equal to zero in~\eqref{eq:stab1}--\eqref{eq:stab5}. Eliminating then $\ph$ and $\wh$ by combining equations~\eqref{eq:stab1}, \eqref{eq:stab3} and \eqref{eq:stab4}, and introducing the transformation 
\begin{align}
    \ut = \uh \tan\phi,
    \quad
    \vt = \vh,
    \quad
    \thetat = \thetah \tan\phi,
    \quad
    \Grast = \Gras \tan\phi \label{transformation}
\end{align}
to remove the explicit dependence on the inclination angle~$\phi$~\cite{Haaland1973}, leads to the three equations
\begin{align}
&\left[{\rm D}(\Theta^\sigma {\rm D}) -(V/\Theta) {\rm D}-(2U-yU')/(4\Theta)  \right]\ut \nonumber \\
  &+\left\{\left[1+VU'+U(2U-y U')/4\right]/\Theta^2 +\sigma \Theta^{\sigma-1} (U' {\rm D}+U'')\right\}\thetat=\Grast  \frac{U'}{\Theta} \vt, \label{eq:vortexeigprob1} \\
    & {\rm D} \left[\frac{1}{l^2} \left({\rm D} \Theta^\sigma -\frac{V}{\Theta} \right) {\rm D}\left({\rm D} -\frac{\Theta'}{\Theta}\right)+{\rm D} \Theta^\sigma -2 \Theta^\sigma \left(2 {\rm D}-\frac{\Theta'}{\Theta}\right) \right] \vt \nonumber \\
    &+\left[\Theta^\sigma {\rm D}\left({\rm D} -\frac{\Theta'}{\Theta} \right) +\frac{1}{\Theta} ({\rm D} V)+ \Theta^\sigma l^2 \right] \vt -\Theta^{-2} \thetat=0, \label{eq:vortexeigprob2} \\
& \left[{\rm D}(\Theta^\sigma {\rm D})+\sigma ({\rm D} \Theta^{\sigma-1} \Theta')- \Pran \frac{V}{\Theta} {\rm D} - \Theta^\sigma l^2 - \Pran \left(\frac{y}{4}U-V\right) \frac{\Theta'}{\Theta^2}\right] \thetat \nonumber \\
& + \Pran \frac{y \Theta'}{4 \Theta} \ut =\Grast  \Pran \frac{\Theta'}{\Theta} \vt, \label{eq:vortexeigprob3} 
\end{align}
with boundary conditions
\begin{align}
    \ut = \vt = {\rm D} \vt = \thetat = 0 \quad \text{at $y = 0$ and as $y \to \infty$}.
    \label{eq:vortexeigprob4}
\end{align}
In the operators acting on the eigenfunctions the symbol ${\rm D}$ denotes differentiation with
respect to $y$. A function of $y$ placed immediately after ${\rm D}$ indicates that
multiplication by that function should be performed prior to differentiation, so that, for
instance, $({\rm D} \Theta^\sigma) \vt={\rm d}(\Theta^\sigma \vt)/{\rm d} y$, while
$(\Theta^\sigma {\rm D})  \vt=\Theta^\sigma {\rm d} \vt/{\rm d} y$ instead. The eigenfunction
$\ut$ is coupled to $\vt$ and $\thetat$ through the last term on the left-hand side in~\eqref{eq:vortexeigprob3}, associated with the slow streamwise variation of the base temperature $\overline{T}$. Also of interest is that for strictly parallel flow (i.e., when the factors listed in~\eqref{factorsnonparallel} are set equal to zero), the problem becomes independent of the base velocity field, and the resulting equations reduce to those corresponding to Rayleigh-B\'enard convection.

\subsection{Numerical results}

\begin{figure}
\centering
\includegraphics[scale=0.5]{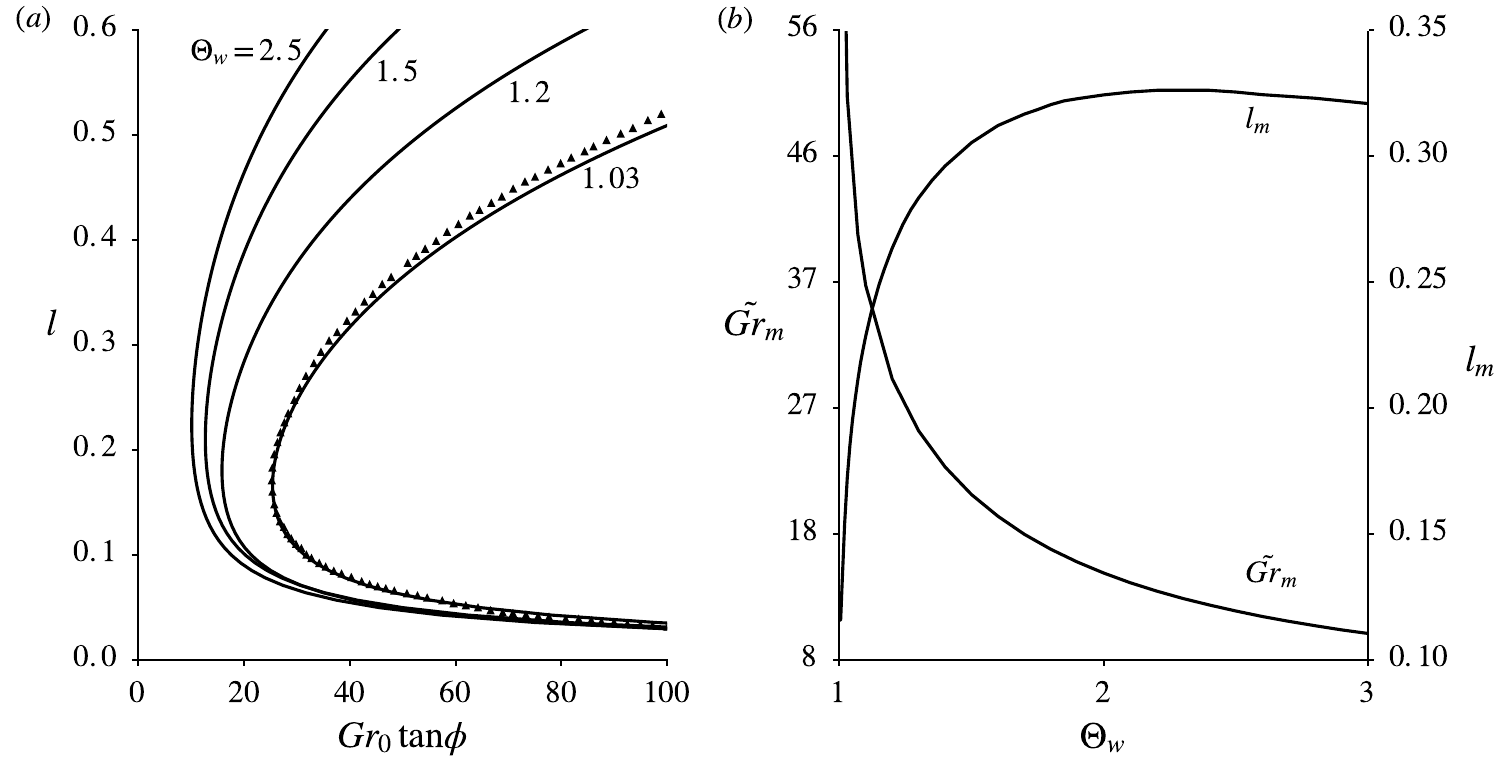}
\caption{%
Results of the non-Boussinesq analysis of the vortex instability, including curves of neutral
stability (a) along with the variation with $\Theta_w$ of $\Grast_m$ and of the associated spanwise wave number
$l_m$ $(b)$; the triangles in $(a)$ corresponds to evaluations of~\eqref{Gw} with the values of
$\Gras_{\text{B}}$ reported in \cite{Haaland1973} (corrected to account for the factor $4$ used
in their definition of the boundary-layer thickness $\deltas$).}
\label{fig:vortexneutral}
\end{figure}

For a given wall-to-ambient temperature ratio $\Theta_w$, which defines the base flow, and a
given spanwise wave number $l$, the above eigenvalue
problem~\eqref{eq:vortexeigprob1}--\eqref{eq:vortexeigprob4} was solved numerically to determine
the critical value $\Grast_0$ of $\Grast$ identifying the conditions of neutral stability. The
variation of $\Grast_0=\tan \phi \, \Gras_0$ with $l$ is shown in
figure~\ref{fig:vortexneutral}$(a)$ for different $\Theta_w$. As can be seen, each neutral curve
$\Grast_0(l)$ reaches a minimum value $\Grast_0=\Grast_m$ at $l=l_m$. This minimum determines
the wave length $2 \pi/l_m$ of the most unstable mode along with the downstream location
$x^*_m=(\Grast_m/\tan \phi)^{4/3} [{\nu_\infty^*}^2/(g^* \cos \phi)]$ at which the vortices
begin to develop. To complete the description, the variation of $\Grast_m$ and $l_m$ for the
most unstable mode is plotted as a function of $\Theta_w$ in
figure~\ref{fig:vortexneutral}$(b)$. It is evident from the plots that increasing the
wall-to-ambient temperature ratio $\Theta_w$ as well as increasing $\phi$ towards a more
horizontal position tends to destabilize the flow. Also noteworthy is the rapid variation of
$\Grast_m$ and $l_m$ with $\Theta_w$ for $\Theta_w-1 \ll 1$.

\subsection{Departures from the Boussinesq approximation}

The results can be compared to the Boussinesq analysis of Haaland \& Sparrow~\cite{Haaland1973} by taking a wall-to-ambient temperature ratio $\Theta_w$ close to unity. The Boussinesq Grashof number employed in~\cite{Haaland1973}, $\Gras_\text{B} = [\betas (\Ts_w - \Ts_\infty) {\xs}^3 \gs \cos \phi/{\nus_\infty}^2]^{1/4}$, is related to that defined in~\eqref{Grasdef} by $\Gras_\text{B} = \Gras [\betas(\Ts_w - \Ts_\infty)]^{1/4}$. The result depends on the temperature used to evaluate the coefficient of thermal expansion $\betas$, given for an ideal gas by $\betas=1/T^*$. Thus, using the ambient temperature to evaluate $\betas_\infty = 1/\Ts_\infty$ yields 
\begin{equation}
\Gras_w=\Gras_{\text{B}} (\Theta_w - 1)^{-1/4}  \label{Gw}
\end{equation}
whereas the expression 
\begin{equation}
\Gras_\infty=\Gras_{\text{B}} (1-\Theta_w^{-1})^{-1/4} \label{Ginfty}
\end{equation}
is obtained by using the wall temperature to evaluate $\betas_w = 1/\Ts_w$. Clearly, the temperature selected to define $\betas$ becomes irrelevant for $\Theta_w - 1 \ll 1$, when the relative differences between both expressions are of order $(\Theta_w-1)/4$. Equation~\eqref{Gw} is selected in the comparison of figure~\ref{fig:vortexneutral}$(a)$ to evaluate the neutral curve for $\Theta_w=1.03$ using the values of $\Gras_\text{B}$ reported in Haaland \& Sparrow~\cite{Haaland1973}, giving the results represented by the triangles.

\begin{figure}
\centering
\includegraphics[scale=0.5]{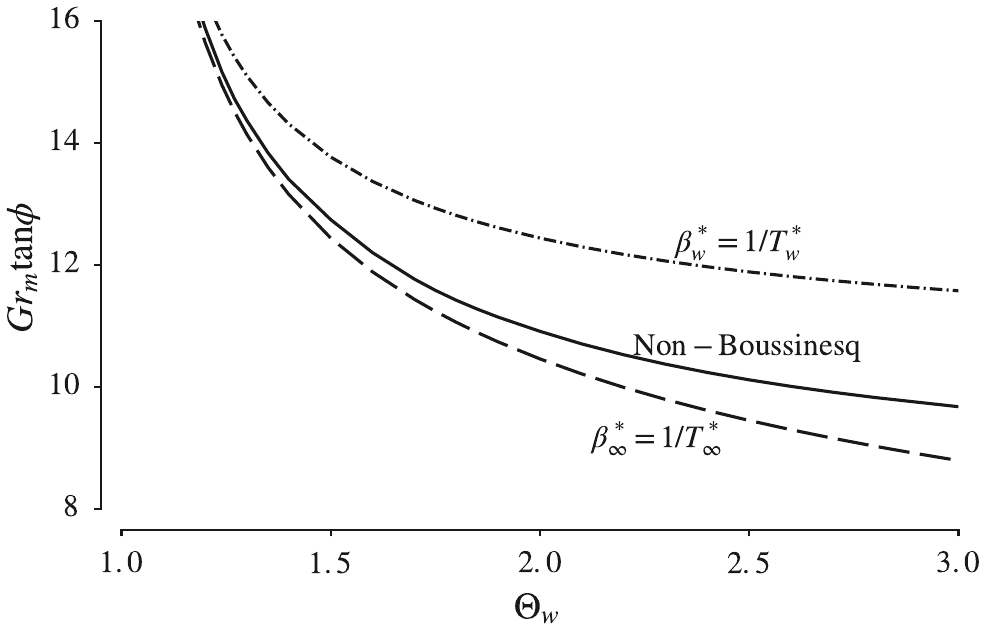}
\caption{The variation of $\Gras_m \tan\phi$ with $\Theta_w$ obtained from integrations of the non-Boussinesq eigenvalue problem~\eqref{eq:vortexeigprob1}--\eqref{eq:vortexeigprob4} (solid curve) and from the Boussinesq  predictions~\eqref{Gw} and~\eqref{Gw} evaluated with $\Gras_{\text{B}}=10.5$, obtained by scaling the value given in \cite{Haaland1973}, as indicated in the text.} \label{fig:vortexBoussinesq}
\end{figure}

As can be seen in figure~\ref{fig:vortexneutral}$(a)$, the differences between the predictions obtained with the Boussinesq approximation and those accounting for non-Boussinesq effects are very small for $\Theta_w=1.03$, with departures in resulting values of $\Grast_m$ and $l_m$ of the order of 1\%. The departures are expected to become larger for increasing values of  $\Theta_w-1$, with the two expressions~\eqref{Gw} and~\eqref{Ginfty} leading to increasingly different values of $\Gras$. This is tested in figure~\ref{fig:vortexBoussinesq} by comparing the curve $\Grast_m-\Theta_w$ obtained from the numerical integrations of~\eqref{eq:vortexeigprob1}--\eqref{eq:vortexeigprob4}, given previously in figure~\ref{fig:vortexneutral}$(b)$, with the predictions derived by extending the Boussinesq approximation with use made of~\eqref{Gw} and~\eqref{Ginfty} evaluated with $\Gras_{\text{B}}=10.5$, the latter obtained by dividing the value $29.6$ reported in~\cite{Haaland1973} by $2\sqrt{2}$, as is needed to account for the factor $4$ used in their definition of the boundary-layer thickness $\delta^*$. As can be seen, our results lie between the two Boussinesq predictions, with the thermal expansion based on the ambient temperature $\betas_\infty = 1/\Ts_\infty$  giving better agreement. For the vortex mode, it is apparent from the curves in the figure that straightforward extension of the stability results computed in the Boussinesq approximation by use of~\eqref{Ginfty} provides sufficiently accurate predictions for the critical Grashof number even for configurations with $\Theta_w-1 \sim 1$. As seen below, the wave mode analyzed next is different in this connection, in that the sharp decrease in $\Grast_m$ found for increasing $\Theta_w$ cannot be predicted by simply extending the results of the Boussinesq approximation, regardless of the temperature used in evaluating the thermal--expansion coefficient.

\section{Wave instability}
\label{sec:ts}

The wave instability on hot inclined plates has received significantly less attention~\cite{HaalandS1973, Iyer1974, Tzuoo1985, Tumin2003} than the related vortex instability problem, possibly because for Boussinesq flow its occurrence its restricted to a moderately small angle range $0^\circ < \phi \lesssim 20^\circ$ about the vertical position. In this connection, it is worth anticipating that the present analysis will show that this range widens significantly for non-Boussinesq gaseous flow with $\Theta_w-1 \sim 1$.

\subsection{The simplified eigenvalue problem}

In accordance with Squire's theorem, only two-dimensional disturbances need to be considered below, so that $l = 0$ in \eqref{eq:normalmodes}. Although a spatial stability analysis is needed to accurately determine the downstream growth rate of the disturbance \cite{Iyer1974, Tumin2003}, the present work focuses on the determination of the curves of neutral stability, associated with real values of both the wave number $k$ and the frequency $\omega$, which can be obtained with a simpler temporal stability analysis. 

With $l=0$ the problem simplifies, in that the solution for $\uh$, $\vh$, $\ph$, and $\thetah$ becomes independent of $\wh$, so that the eigenvalue problem reduces to the solution of ~\eqref{eq:stab1}--\eqref{eq:stab3} and~\eqref{eq:stab5} with the homogeneous boundary conditions
\begin{align}
    \uh = \vh = \thetah = 0 \text{ at $y=0$,}
    \quad \text{and} \quad
    \uh, \vh, \thetah, \ph \rightarrow 0 \text{ as $y\to\infty$.}
\end{align}
The eigenvalue nature of this problem becomes more apparent by casting the equations in the standard form $A \vect{f} = \omega B \vect{f}$, with $A$ and $B$ being linear differential operators acting on $\vect{f} = (\uh, \vh, \ph, \thetah)$. As a result of the non-Boussinesq character of the continuity equation, elimination of $\ph$ and $\uh$ by linear combinations leads to a quadratic eigenvalue problem for the eigenvalue $\omega$, which is not further considered here because of the additional complications involved in its numerical solution.

\subsection{Critical curves of neutral stability}

\begin{figure}
\centering
\includegraphics[scale=0.5]{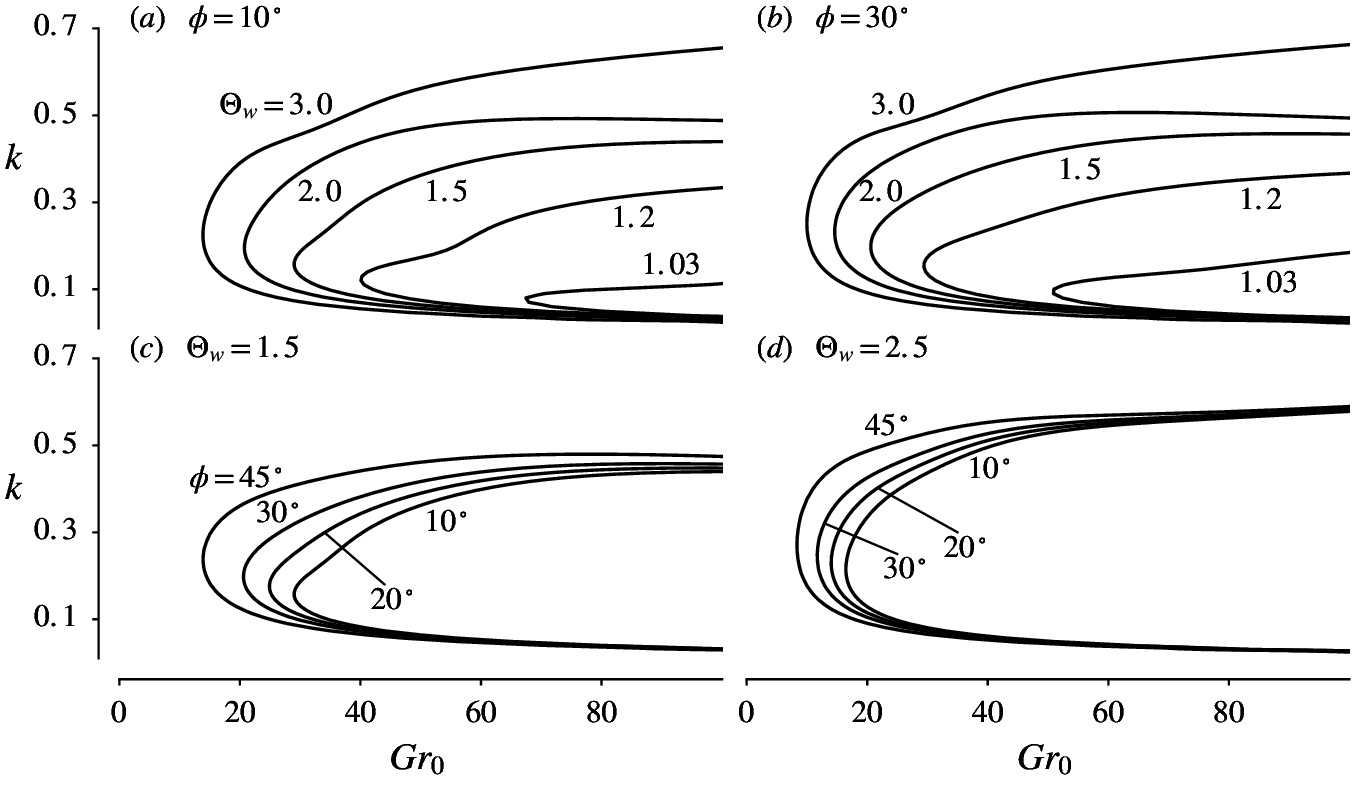}
\caption{Curves of neutral stability for the wave mode with different wall-to-ambient temperature ratios $\Theta_w$ and different inclination angles $\phi$.
}
\label{fig:wave}
\end{figure}

Results obtained by integrating numerically the eigenvalue problem for the wave instability are shown in figure~\ref{fig:wave}. Unlike the vortex instability, for which the dependence on the inclination angle can be scaled out of the equations by the transformation~\eqref{transformation}, thereby reducing the parametric dependence of the results, the eigenvalue problem for the wave instability depends on both $\phi$ and $\Theta_w$. Curves of neutral stability are plotted in figures~\ref{fig:wave}$(a)$ and~\ref{fig:wave}$(b)$ for a fixed inclination angle $\phi$ and different values of the wall temperature ratio $\Theta_w$, with corresponding curves for fixed $\Theta_w$ and different values of $\phi$ given in figures~\ref{fig:wave}$(c)$ and~\ref{fig:wave}$(d)$. Just as for the vortex instability, increasing the wall-to-ambient temperature ratio, as well as increasing the inclination angle (a more horizontal plate) has a destabilizing effect on the flow. The rapid change in the neutral curves shown in the upper plots of the figure as the wall temperature increases from $\Theta_w=1.03$ to $\Theta_w=1.2$ is indicative of the importance of non-Boussinesq effects for wave instability. 

\subsection{Stability predictions based on vorticity--stream function formulation}

\begin{figure}
\centering
\includegraphics[scale=0.5]{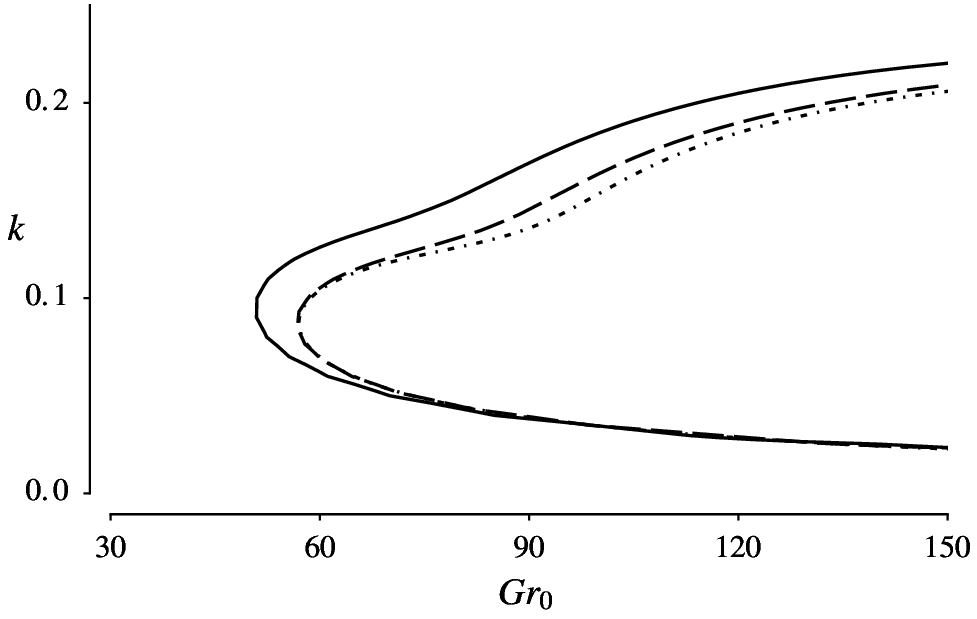}
\caption{Curves of neutral stability for the wave-instability mode with $\Theta_w = 1.03$ and $\phi = 30^\circ$ obtained from \eqref{eq:stab1}, \eqref{eq:stab5} and~\eqref{eqcomp} (solid curve) and by repeating the computations with the underlined term removed in~\eqref{eqcomp} (dashed curve); the dotted curve corresponds to evaluations of~\eqref{Gw} with the values of $\Gras_{\text{B}}$ reported in \cite{HaalandS1973}.}
\label{fig:wavehaaland}
\end{figure}

Before continuing the discussion of results, it is worth commenting on the differences arising
from the use of different starting conservation equations in the presence of a slowly varying
base flow. Unlike many previous analyses, which were derived on the basis of a vorticity--stream
function formulation, the present analysis employs as a starting point the conservation
equations written in primitive variables. Although the logical expectation is that both
approaches yield the same results, that being the case for strictly parallel base flow, small
but noticeable discrepancies were found between both computational approaches when the
nonparallel terms are retained for the base flow. This is so because when the pressure is
eliminated from the equations written in primitive variables \emph{after} the normal mode
decomposition is introduced, \ie from equations~\eqref{eq:stab1}--\eqref{eq:stab5}, to obtain a
normal-mode vorticity form, the resulting set of equations is not identical to that obtained
by writing the momentum equation in terms of the vorticity \emph{before} introducing the normal
mode decomposition. This discrepancy does not happen when analyzing the vortex mode, because the
direction of the vorticity differs between the base flow and the perturbations, i.e. the
base-flow vorticity is oriented along the spanwise direction while the perturbed vorticity is
oriented along the streamwise direction. 

These differences can be illustrated by comparing the results of our formulation for $\Theta_w
\ll 1$ with those obtained earlier by Haaland \& Sparrow \cite{HaalandS1973} using the
linearized vorticity--stream function Boussinesq formalism presented earlier in
\cite{Haaland1973stability}. The comparison can be readily established by writing our equations
\eqref{eq:stab2} and~\eqref{eq:stab3} for $\Theta_w-1 \ll 1$ followed by elimination of the
pressure to yield
\begin{align}
&[{\rm D}^3-V {\rm D}^2- \Gras \, \ui (Uk-\omega) {\rm D} -k^2 {\rm D}+ V''] \uh \nonumber \\
&[-\ui k {\rm D}^2 +\ui k V {\rm D} - \Gras \, k (U k-\omega) -\Gras U'' + \ui k^3+\underline{\ui k V'}] \vh +[{\rm D} -\ui k \tan \phi] \thetah=0 \label{eqcomp}
\end{align}
after the continuity equation is used to simplify the result. Comparing the above equation with
equation (15) in \cite{Haaland1973stability} reveals that both are identical, except for the
underlined term in~\eqref{eqcomp}, which is absent in \cite{Haaland1973stability}, that being a
consequence of the different derivation procedure. As previously mentioned, the missing term,
proportional to the transverse gradient of the transverse base velocity, would be absent for
strictly parallel flow.  Its quantitative importance is assessed in
figure~\ref{fig:wavehaaland}, where the neutral curve obtained using \eqref{eq:stab1}
and~\eqref{eq:stab5} together with~\eqref{eqcomp} (solid curve) is compared with that obtained
after selectively removing the term $\ui k V'$ in~\eqref{eqcomp} (dashed curve). The additional
term is seen to destabilize the flow, resulting in critical Grashof numbers that are about 10\%
smaller than those obtained on the basis of the vorticity--stream function formulation. As
expected, the dashed curve is seen to agree with that obtained by evaluating~\eqref{Gw} with the
values of $\Gras_{\text{B}}$ reported in Haaland \& Sparrow \cite{HaalandS1973} for
$\phi=30^{\rm o}$, represented by a dotted line in figure~\ref{fig:wavehaaland}.

\subsection{Quantitative assessment of non-Boussinesq effects}

\begin{figure}
\centering
\includegraphics[scale=0.5]{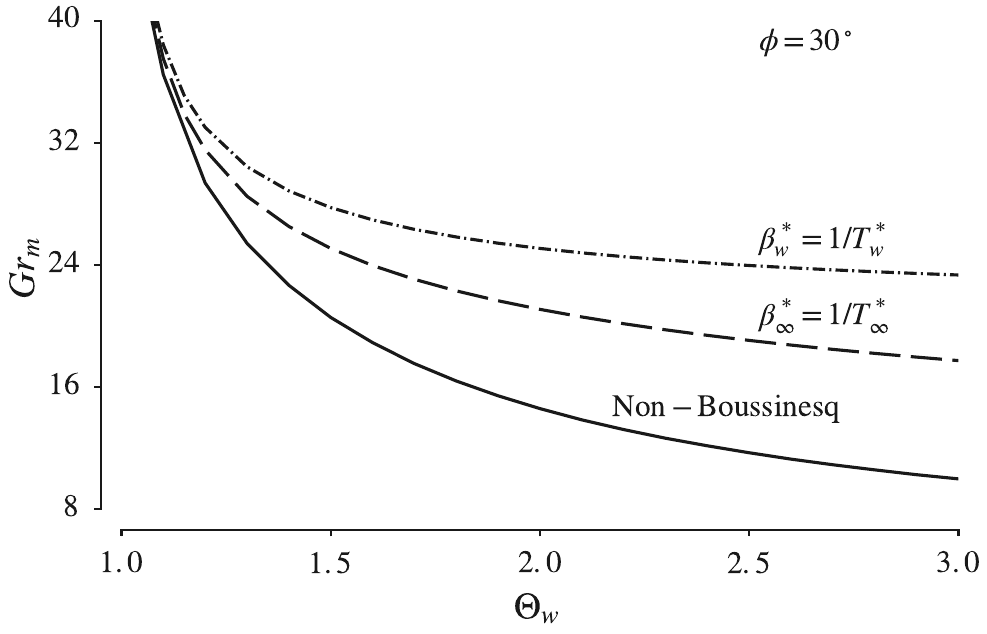}
\caption{%
The variation of $\Gras_m$ with $\Theta_w$ for $\phi=30^\circ$ obtained for the wave instability
from integrations of the non-Boussinesq eigenvalue problem (solid curve) and from the Boussinesq
predictions~\eqref{Gw} and~\eqref{Ginfty} evaluated with $\Gras_{\text{B}}=17.96$, the latter
computed by extrapolating the numerical results as $\Theta_w \rightarrow 1$.}
\label{fig:waveBoussinesq}
\end{figure}

To complete the description, the Boussinesq  predictions~\eqref{Gw} and~\eqref{Ginfty} are
compared in figure~\ref{fig:waveBoussinesq} with the results of the non-Boussinesq eigenvalue
problem for $\phi=30^{\rm o}$. The value $\Gras_{\text{B}}=21.11$ used in the evaluations
of~\eqref{Gw} and~\eqref{Ginfty} is extracted from our numerical results for $\Theta_w-1 \ll 1$.
Because of the discrepancies illustrated in figure~\ref{fig:wavehaaland}, this value is somewhat
smaller than the value reported in \cite{HaalandS1973}, which gives $\Gras_{\text{B}}=23.63$
once corrected to account for differences in their definition of $\delta^*$. As can be seen in
figure~\ref{fig:waveBoussinesq}, the extrapolations of the Boussinesq results consistently
overpredict the critical Grashof number. This is so because the Boussinesq approximation does
not describe accurately the acceleration of the light gas for increasing $\Theta_w$, which is
apparent in the velocity profiles in figure~\ref{fig:baseflow}, resulting in augmented shear
stresses that trigger the wave--mode instability. 

\section{Variation of the crossover angle}

\begin{figure}
\centering
\includegraphics[scale=0.5]{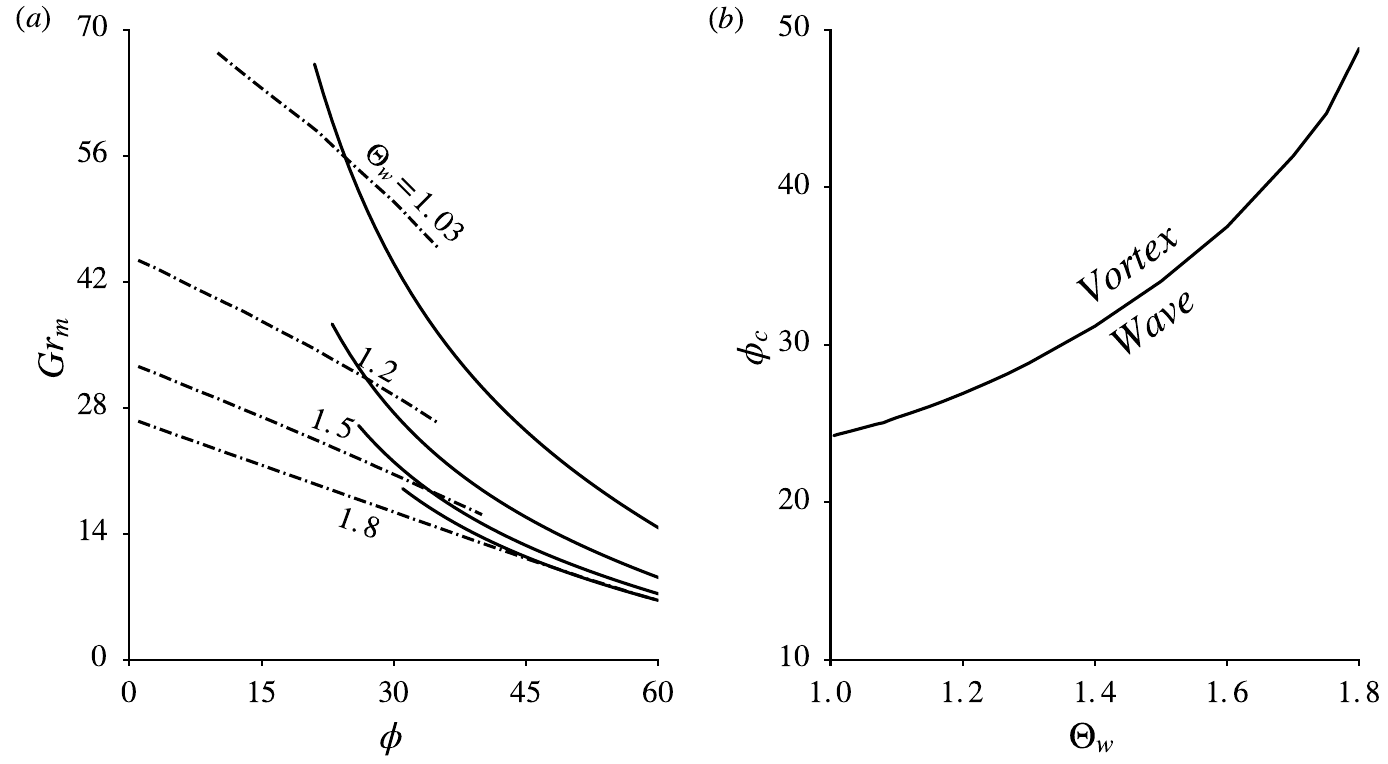} 
\caption{The left-hand-side plot represents the variation of the critical Grashof number with the inclination angle for the wave mode (dashed curves) and for the vortex mode (solid curves) for different values of the wall-to-ambient temperature ratio $\Theta_w$, while the right-hand-side plot gives the variation with $\Theta_w$ of the crossover angle $\phi_c$ at which each pair of curves crosses.} \label{fig:crossover}
\end{figure}

It is apparent from the comparison of the curves in figures~\ref{fig:vortexBoussinesq} and~\ref{fig:waveBoussinesq} that non-Boussinesq effects have a larger influence on the wave mode than they have on the vortex mode. For both modes, increasing the plate temperature promotes instability by decreasing the associated critical Grashof number, but the decrease of $\Gras_m$ is much more pronounced for the wave mode, thereby suggesting that the character of the instability that develops over a plate at a given inclination angle may switch from a vortex mode to a wave mode as the plate temperature increases. 

To investigate this aspect of the problem we compare in figure~\ref{fig:crossover}$(a)$ the curves $\Gras_m=\Grast_m/\tan \phi$ corresponding to the vortex mode, with $\Grast_m(\Theta_w)$ given in figure~\ref{fig:vortexneutral}$(b)$, with the values of $\Gras_m$ obtained from the turning points of the neutral curves computed for the wave mode for a fixed value of $\Theta_w$, given for instance in figure~\ref{fig:wave}$(c)$ for the particular case $\Theta_w=1.5$. The crossing point of each pair of curves determines the angle $\phi_c$ at which the character of the most unstable mode changes from a traveling wave (for $\phi < \phi_c$) to stationary streamwise vortices (for $\phi > \phi_c$). The resulting crossover angle increases for increasing $\Theta_w$, as shown in figure~\ref{fig:crossover}$(b)$, which delineates the parametric regions where each one of the modes prevails. 

Our linear stability analysis predicts a crossover angle $\phi_c=23.8^\circ$ for $\Theta_w-1 \ll 1$, in agreement with the value $\phi_c=22^\circ$ reported in previous linear stability analyses \cite{Chen1982,Tzuoo1985} using instead the scalings of the horizontal boundary layer for the self-similar base flow. A somewhat smaller value was estimated in the work of Kahawita and Meloney \cite{Kahawita1974} by evaluating approximately the critical Grashof number of the wave mode from the results of the vertical plate through simple multiplication by $\cos \phi$. It is worth noting that the values observed in experiments are typically somewhat smaller than the theoretical predictions (e.g., values in the range $14^{\circ} < \phi_c < 17^{\circ}$ are reported in the early work~\cite{Lloyd1970}). The deviations are attributable to the limitations of the linear analysis performed here, although more work is needed to clarify this issue. 

As the value of $\Theta_w$ increases the crossing points of the curves in figure~\ref{fig:crossover}$(a)$ moves to the right, corresponding a wider range $0\le \phi < \phi_c$ where the wave mode prevails. Unexpectedly, the two curves $\Gras_m-\phi$ become tangent at $\Theta_w \simeq 1.8$ and exhibit no crossing point for larger values of $\Theta_w >1.8$, for which our analysis predicts the wave mode to be dominant for all angles. Clearly, however, this prediction would need to be reconsidered for near-horizontal plates with values of $\phi$ approaching $\pi/2$, for which the condition~\eqref{condition}, used in deriving the governing equations for the inclined plate, would no longer hold. A revised stability analysis, accounting for transverse pressure differences across the boundary layer in deriving the governing equations for both the base flow and its perturbations, would be needed to assess the transition between both instability modes for very hot plates with $\Theta_w > 1.8$.

\section{Concluding remarks}
\label{sec:discconc}

Although the existing experimental evidence and the previous theoretical predictions based on
the Boussinesq approximation indicate that at higher wall temperatures the flow tends to become
unstable at lower values of the Grashof number, corresponding to smaller downstream distances
from the plate edge, quantitative results for relative wall-to-ambient temperature differences
of order unity are not currently available. This paper provides the needed quantification on the
basis of a non-Boussinesq temporal stability analysis accounting for the slow variation of the
base flow. The results provide predictions of critical Grashof numbers for both modes of
instability and their associated wave numbers. While non-Boussinesq effects are found to have
only a moderate quantitative effect on the instability mode involving streamwise vortices, it is
found that the augmented shear resulting from the flow acceleration in the presence of large
density differences promotes wave instabilities significantly. As a result, the range of angles
about the vertical position where the wave mode is dominant is predicted to increase
substantially with increasing wall temperatures, a finding of our linear stability analysis that
warrants future experimental investigation.

The characteristics of the analysis presented, both linear and local, limit the accuracy of some
of the predictions. For instance, the nonlinear growth of the perturbations, not described
herein, may explain the discrepancies between the predicted value of $\phi_c=23.8^\circ$  for
$\Theta_w-1 \ll 1$ and previous experimental observations. Also, although
the analysis accounts for the slow streamwise variation of the base flow
in~\eqref{decomposition}, the shape functions of the accompanying perturbations, as well as
the wave numbers $k$ and $l$ are not allowed to depend on the rescaled coordinate $x/\Gras$,
as would be needed to account for their downstream evolution. Incorporating the latter,
while still taking advantage of the slenderness of the boundary-layer flow associated with
moderately large values of the Grashof number---thus retaining only terms up to
$\mathcal{O}(\Gras^{-1})$---would turn the \emph{local} stability problem into a
\emph{nonlocal} parabolic stability problem that has to be integrated in the downstream
direction.
In this type of treatment a unique neutral curve cannot be defined since the
streamwise development depends strongly on the initial conditions that are imposed at a certain
location, as pointed out for the first time by Hall~\cite{Hall1983} for the related problem of
G\"ortler instability over a curved wall.
Subsequent studies~\cite{bottaro1999} have found that, although different initial conditions
result in different growth rates during the transient stage, at very large downstream
distances the local growth rate is found to be independent of the initially imposed disturbance.
Although an attempt to apply these concepts to the present problem was made by
Tumin~\cite{Tumin2003}, more work is clearly needed to further clarify the instability
characteristics of natural-convection flows.

A global stability analysis of the natural-convection flow under consideration, in which the
instabilities are considered as two-dimensional temporal Fourier modes, is also worth pursuing
in future work.  Since this alternative approach does not invoke the slenderness condition of
the basic flow, it might prove useful to study cases with very large wall-to-ambient temperature
ratios, for which, as shown in the present work, the instability sets in at smaller Grashof
numbers, and thus at shorter distances from the leading edge. The highly nonparallel
Navier-Stokes region close the edge of the plate then may play an important role in these early
transition events.

\section*{Acknowledgements}

We would like to thank one of the reviewers for constructive comments
and suggestions for relevant references.

\appendix

\section{Stability equations}
\label{sec:appstabeqs}

Introducing the normal-mode decomposition~\eqref{eq:normalmodes} into~\eqref{eq:lin1}--\eqref{eq:lin4} with $\rhoh = -\Theta^{-2} \thetah$ and $\muh = \sigma \Theta^{\sigma-1} \thetah$ and discarding terms of order $\Gras^{-2}$ yields
\begin{align}
    & \left[\Gras \, \ui k \Theta^{-1}+(y/4) \Theta^{-2}\Theta' \right] \uh +\Gras ({\rm D}\Theta^{-1}) \vh+ \Gras \, \ui \, l \Theta^{-1} \wh \nonumber \\ 
    &-  \left[{\rm D}V +\Gras \, \ui \,(Uk-\omega)+(2U-y U')/4\right] \Theta^{-2} \thetah = 0, \label{eq:stab1}\\   
    &[{\rm D}(\Theta^\sigma {\rm D})- \Theta^{-1} V {\rm D} - \Gras \, \ui \, \Theta^{-1} (U k - \omega) - \Theta^{-1}(2U-yU')/4  \nonumber \\ 
    & -\Theta^{\sigma}(2 k^2 + l^2)] \uh+\left[\ui k ({\rm D} \Theta^\sigma)-\Gras \Theta^{-1} U'\right] \vh - kl \Theta^{\sigma} \wh - \Gras \, \ui \, k \ph \nonumber \\ 
    &+[\sigma ({\rm D} U' \Theta^{\sigma+1})+U(2U-yU')/4+V U'+1] \Theta^{-2} \thetah=0, \label{eq:stab2} \\
    & (\ui k \Theta^\sigma {\rm D}) \uh +\left[2{\rm D}(\Theta^\sigma {\rm D})-\Theta^{-1} ({\rm D} V) -\Gras \, \ui \,  \Theta^{-1} (Uk -\omega)-\Theta^{\sigma}  (k^2 + l^2) \right] \vh\nonumber \\ 
    &+(\ui \, l \, \Theta^\sigma {\rm D}) \wh - \Gras \,  {\rm D} \ph +(\tan \phi \, \Theta^{-2}+ \sigma \, \ui \, k \, U' \Theta^{\sigma-1}) \thetah=0, \label{eq:stab3}\\
    &- \Theta^\sigma k \, l \, \uh +\ui \,  l  ({\rm D} \Theta^\sigma) \vh + [{\rm D} (\Theta^\sigma {\rm D})-\Theta^{-1} V {\rm D}\nonumber \\  &-\Gras \, \ui \Theta^{-1} (U k - \omega)-\Theta^{\sigma} (k^2 + 2 l^2)  ] \wh  - \Gras \,  \ui l \ph=0, \label{eq:stab4}\\
    & (\Pran \Theta^{-1} \Theta' y/4) \uh - \Gras \Pran \Theta^{-1} \Theta' \vh + [{\rm D} (\Theta^\sigma {\rm D} +\sigma \Theta' \Theta^{\sigma-1}) -\Pran \Theta^{-1} V {\rm D} \nonumber \\
    &  -\Gras \Pran \, \ui \Theta^{-1} (Uk-\omega)-\Theta^{\sigma} (k^2+l^2) +\Pran (V - U y/4) \Theta^{-2}\Theta'] \thetah=0, \label{eq:stab5} 
            \end{align}
        to be integrated with boundary conditions 
        \begin{align}
    \uh = \vh = \wh =\thetah = 0 \text{ at $y=0$,}
    \quad \text{and} \quad
    \uh=\vh=\wh=\thetah=\ph=0 \text{ as $y\to\infty$.}
\end{align}
As in the main text, the symbol ${\rm D}$ denotes differentiation with respect to $y$. In the convention adopted a function of $y$ placed after ${\rm D}$ indicates that multiplication by that function should be performed prior to differentiation.

\section*{References}

\bibliography{references}

\begin{thebibliography}{10}
\expandafter\ifx\csname url\endcsname\relax
  \def\url#1{\texttt{#1}}\fi
\expandafter\ifx\csname urlprefix\endcsname\relax\def\urlprefix{URL }\fi
\expandafter\ifx\csname href\endcsname\relax
  \def\href#1#2{#2} \def\path#1{#1}\fi

\bibitem{SB1930}
E.~Schmidt, N.~Beckmann, Das temperature and geschwindigkeitsfeld von einer
  warmeabgebenden senkrechten platte bei naturlicher konvecktion, Tech. Mech.
  Thermodynamic 1 (1930) 341--349, 391--406.

\bibitem{Sparrow1969}
E.~Sparrow, R.~Husar, Longitudinal vortices in natural convection flow on
  inclined plates, J.~Fluid Mech. 37~(02) (1969) 251--255.

\bibitem{Haaland1973}
S.~Haaland, E.~Sparrow, Vortex instability of natural convection flow on
  inclined surfaces, Int.~J.~Heat Mass Transfer. 16~(12) (1973) 2355--2367.

\bibitem{HaalandS1973}
S.~Haaland, E.~Sparrow, Wave instability of natural convection on inclined
  surfaces accounting for nonparallelism of the basic flow, J.~Heat Transfer.
  95~(3) (1973) 405--407.

\bibitem{Lloyd1970}
J.~Lloyd, E.~Sparrow, On the instability of natural convection flow on inclined
  plates, J.~Fluid Mech. 42~(03) (1970) 465--470.

\bibitem{Gebhart1978}
H.~Shaukatullah, B.~Gebhart, An experimental investigation of natural
  convection flow on an inclined surface, Int.~J.~Heat Mass Transfer. 21~(12)
  (1978) 1481--1490.

\bibitem{Cheng1988}
K.~Cheng, Y.~Kim, Flow visualization studies on vortex instability of natural
  convection flow over horizontal and slightly inclined constant-temperature
  plates, J.~Heat Transfer. 110~(3) (1988) 608--615.

\bibitem{Zuercher1998}
E.~Zuercher, J.~Jacobs, C.~Chen, Experimental study of the stability of
  boundary-layer flow along a heated, inclined plate, J.~Fluid Mech. 367 (1998)
  1--25.

\bibitem{Jeschke2000}
P.~Jeschke, H.~Beer, Longitudinal vortices in a laminar natural convection
  boundary layer flow on an inclined flat plate and their influence on heat
  transfer, J.~Fluid Mech. 432 (2001) 313--339.

\bibitem{Trautman2002}
M.~Trautman, A.~Glezer, The manipulation of the streamwise vortex instability
  in a natural convection boundary layer along a heated inclined flat plate,
  J.~Fluid Mech. 470 (2002) 31--61.

\bibitem{Kimura2003}
F.~Kimura, K.~Kitamura, M.~Yamaguchi, T.~Asami, Fluid flow and heat transfer of
  natural convection adjacent to upward-facing, inclined, heated plates, Heat
  Transfer - Asian Research. 32~(3) (2003) 278--291.

\bibitem{Hwang1973}
G.~Hwang, K.~Cheng, Thermal instability of laminar natural convection flow on
  inclined isothermal plates, The Canadian J.~Chemical Eng. 51~(6) (1973)
  659--666.

\bibitem{Kahawita1974}
R.~Kahawita, R.~Meroney, The vortex mode of instability in natural convection
  flow along inclined plates, Int.~J.~Heat Mass Transfer. 17~(5) (1974)
  541--548.

\bibitem{Iyer1974}
P.~Iyer, R.~Kelly, The stability of the laminar free convection flow induced by
  a heated inclined plate, Int.~J.~Heat Mass Transfer. 17~(4) (1974) 517--525.

\bibitem{Chen1982}
T.~Chen, K.~Tzuoo, Vortex instability of free convection flow over horizontal
  and inclined surfaces, J.~Heat Transfer. 104~(4) (1982) 637--643.

\bibitem{Tzuoo1985}
K.~L. Tzuoo, T.~S. Chen, B.~F. Armaly, Wave instability of natural convection
  flow on inclined surfaces, J.~Heat Transfer. 107~(1) (1985) 107--111.

\bibitem{Lin2001}
M.~Lin, Numerical study of formation of longitudinal vortices in natural
  convection flow over horizontal and inclined surfaces, Int.~J.~Heat Mass
  Transfer. 44~(9) (2001) 1759--1766.

\bibitem{Tien1986}
H.~Tien, T.~Chen, B.~Armaly, Vortex instability of natural convection flow over
  horizontal and inclined plates with uniform surface heat flux, Numerical Heat
  Transfer. 9~(6) (1986) 697--713.

\bibitem{Chen1991}
C.~Chen, A.~Labhabi, H.~Chang, R.~Kelly, Spanwise pairing of finite-amplitude
  longitudinal vortex rolls in inclined free-convection boundary layers,
  J.~Fluid Mech. 231 (1991) 73--111.

\bibitem{Tumin2003}
A.~Tumin, The spatial stability of natural convection flow on inclined plates,
  J.~Fluids Eng. 125~(3) (2003) 428--437.

\bibitem{Saha2011}
S.~Saha, J.~Patterson, C.~Lei, Scaling of natural convection of an inclined
  flat plate: sudden cooling condition, J. Heat Transfer 133 (2011) 041503.

\bibitem{Dou2013}
H.-S. Dou, G.~Jiang, C.~Lei, Numerical simulation and stability study of
  natural convection in an inclined rectangular cavity, Math. Probl. Eng. 2013
  (2013) 1--12.

\bibitem{Clarke1975}
J.~Clarke, N.~Riley, Natural convection induced in a gas by the presence of a
  hot porous horizontal surface, J.~Mechanics and Applied Mathematics. 28~(4)
  (1975) 373--396.

\bibitem{Ackroyd1976}
J.~Ackroyd, Laminar natural convection boundary layers on near-horizontal
  plates, Proc. R. Soc. Lond. 352~(1669) (1976) 249--274.

\bibitem{Sanchez2013}
M.~Gollner, A.~S{\'a}nchez, F.~Williams, On the heat transferred to the air
  surrounding a semi-infinite inclined hot plate, J.~Fluid Mech. 732 (2013)
  304--315.

\bibitem{schmid2012}
P.~J. Schmid, D.~S. Henningson, Stability and transition in shear flows, Vol.
  142, Springer Science \& Business Media, 2012.

\bibitem{Haaland1973stability}
S.~Haaland, E.~Sparrow, Stability of buoyant boundary layers and plumes, taking
  account of nonparallelism of the basic flows, Journal of Heat Transfer 95~(3)
  (1973) 295--301.

\bibitem{Hall1983}
P.~Hall, The linear development of g\"ortler vortices in growing boundary
  layers, J.~Fluid Mech. 130 (1983) 41--58.

\bibitem{bottaro1999}
A.~Bottaro, P.~Luchini, G{\"o}rtler vortices: Are they amenable to local
  eigenvalue analysis?, Eur.~J.~Mech. B/Fluids 18~(1) (1999) 47--65.

\end{thebibliography}

\end{document}